\documentclass[aps,prd,twocolumn,groupedaddress,floatfix,nofootinbib,notitlepage]{revtex4-1}

\usepackage[a4paper, total={6.83in, 9.47in}]{geometry}

\usepackage[utf8x]{inputenc}

\usepackage{graphicx}
\usepackage{hyperref}
\usepackage{latexsym}
\usepackage{amsmath}
\usepackage{amssymb}
\usepackage{pdfsync}
\usepackage{subfigure}
\usepackage{slashed}

\newcommand{\be}{\begin{equation}}
\newcommand{\ee}{\end{equation}}
\newcommand{\bea}{\begin{equation}\begin{aligned}}
\newcommand{\eea}{\end{aligned}\end{equation}}
\def\lsim{\mathrel{\raise.3ex\hbox{$<$\kern-.75em\lower1ex\hbox{$\sim$}}}}
\def\gsim{\mathrel{\raise.3ex\hbox{$>$\kern-.75em\lower1ex\hbox{$\sim$}}}}

\begin{document}

\title{Electroweak baryogenesis and gravitational waves from a real scalar singlet}

\author{Ville Vaskonen} \email{ville.vaskonen@kbfi.ee}
\affiliation{National Institute of Chemical Physics and Biophysics, \\ R\"avala 10, 10143 Tallinn, Estonia}

\begin{abstract}
We consider a real scalar singlet field which provides a strong first-order electroweak phase transition via its coupling to the Higgs boson, and gives a $CP$ violating contribution on the top quark mass via a dimension-6 operator. We study the correlation between the baryon-to-entropy ratio produced by electroweak baryogenesis, and the gravitational wave signal from the electroweak phase transition. We show that future gravitational wave experiments can test, in particular, the region of the model parameter space where the observed baryon-to-entropy ratio can be obtained even if the new physics scale, which is explicit in the dimension-6 operator, is high.
\end{abstract}

\maketitle


\section{Introduction}
Electroweak baryogenesis is one of the most studied scenarios for dynamically producing the matter-antimatter asymmetry in the Universe \cite{Kuzmin:1985mm,Cohen:1990py}. The scenario relies on a strong first-order electroweak phase transition during which the baryon number violating sphaleron processes translate the $CP$ asymmetry at the bubble wall region to baryon asymmetry. In the Standard Model the electroweak phase transition is a crossover \cite{Kajantie:1996mn,Aoki:1999fi}, and the $CP$ violating phase of the CKM matrix is generally agreed to be too weak to account for the observed baryon-to-entropy ratio \cite{Jarlskog:1985ht,Farrar:1993hn,Gavela:1993ts,Konstandin:2003dx}. New physics can, however, modify the electroweak phase transition and provide new sources of $CP$ violation, possibly enabling a successful electroweak baryogenesis.

Probably the simplest model in which a strong first-order electroweak phase transition can be realized is the real scalar singlet extension of the Standard Model \cite{McDonald:1993ey,Espinosa:1993bs,Choi:1993cv,Ham:2004cf,Espinosa:2007qk,Profumo:2007wc,Ahriche:2007jp,Ahriche:2012ei,Alanne:2014bra,Profumo:2014opa,Alanne:2016wtx,Tenkanen:2016idg}. The model can be extended by introducing an effective nonrenormalizable coupling between the top quark and the singlet scalar, which modifies the top quark mass at nonzero values of the singlet scalar field \cite{Espinosa:2011eu,Cline:2012hg}. If this coupling is complex, it provides a source of $CP$ violation, thereby making electroweak baryogenesis in this scenario possible.

Another interesting aspect of first-order phase transitions is that they produce gravitational waves \cite{Steinhardt:1981ct,Hogan:1984hx,Witten:1984rs}, which can perhaps be observed in future space-based gravitational wave interferometers \cite{Corbin:2005ny,Seoane:2013qna,Caprini:2015zlo}. Gravitational wave signals from a first-order electroweak phase transition have recently been extensively studied in various extensions of the Standard Model \cite{Kakizaki:2015wua,Huber:2015znp,Leitao:2015fmj,Chala:2016ykx,Huang:2016cjm,Hashino:2016xoj,Artymowski:2016tme,Huang:2016odd}. Also, different scenarios with gravitational waves from hidden sector phase transitions and from phase transitions at energy scales above the electroweak transition have been considered \cite{Schwaller:2015tja,Jinno:2015doa,Jaeckel:2016jlh,Dev:2016feu,Jinno:2016knw,Katz:2016adq,Kubo:2016kpb}. Yet, the correlation between the gravitational wave signal from the electroweak phase transition and the baryon-to-entropy ratio produced by electroweak baryogenesis has not been studied. 

As shown in Ref.~\cite{No:2011fi}, a sizable gravitational wave signal can be obtained while simultaneously satisfying the requirements for viable electroweak baryogenesis. The gravitational wave signal is strongest for high bubble wall velocities, whereas the baryon-to-entropy ratio produced via electroweak baryogenesis decreases as a function of the relevant velocity. However, for electroweak baryogenesis the relevant velocity is not directly the bubble wall velocity, but the relative velocity between the bubble wall and the plasma just in front of the wall. Especially for very strong transitions this velocity is much lower than the bubble wall velocity.

In the real scalar singlet extension of the Standard Model a first-order electroweak phase transition can be realized already at tree-level by a two-step transition pattern where first the singlet scalar obtains a nonzero vacuum expectation value. This phase transition pattern can lead to strong supercooling. As the transition finally happens, a large amount of vacuum energy is released, so the gravitational wave signal from the transition can be strong. In this paper we show that the baryon-to-entropy ratio produced by electroweak baryogenesis, and the gravitational wave signal from electroweak phase transition are correlated in the real scalar singlet extension of the Standard Model. We compare the gravitational wave signal to the expected sensitivities of Laser Interferometer Space Antenna (LISA)~\cite{Baker:2007} and Big Bang Observer (BBO)~\cite{Phinney:2004}, and show that these experiments can test the model.

The paper is organized as follows: First, in Sec.~\ref{model} we introduce the model, and in Sec.~\ref{ptpattern} we discuss the phase transition pattern giving a strong first-order electroweak phase transition. Then, in Sec.~\ref{nucleation} we calculate the bubble nucleation temperature. In Sec.~\ref{ewbg} we perform the electroweak baryogenesis calculation by solving the transport equations, and we study the dependence of the baryon-to-entropy ratio on bubble wall velocity and width. In Sec.~\ref{GW} we calculate the gravitational wave signal produced by the electroweak phase transition and compare it to the expected sensitivities of future gravitational wave interferometers. Finally, in Sec.~\ref{conclusions} we present our conclusions.

\section{Model} \label{model}
We consider the simplest scalar extension of the Standard Model where, in addition to the Standard Model Higgs doublet $H$, the scalar sector includes a $Z_2$ symmetric real scalar singlet field $s$. The scalar potential of the model is given by
\bea
\label{scalarpotential}
V(H,s) =& \mu_{\rm h}^2 H^{\dagger}H + \lambda_{\rm h}(H^{\dagger}H)^2 \\
&+ \frac{\lambda_{\rm hs}}{2}(H^{\dagger}H)s^2  + \frac{\mu_{\rm s}^2}{2}s^2 + \frac{\lambda_{\rm s}}{4}s^4.
\eea
As we will explain in the next section, we consider a phase transition pattern where at $T=0$ the vacuum expectation value of $s$ is zero. The Higgs field mass term is related to the $T=0$ vacuum expectation value of the Higgs field, $v=246$ GeV, via $\mu_{\rm h}^2 = -\lambda_{\rm h} v^2$, and the results from LHC \cite{Aad:2015zhl} fix the mass of the Higgs boson, $m_{\rm h}^2 = 2\lambda_{\rm h} v^2$, to $m_{\rm h} = 125$ GeV. The scalar potential then includes only three free parameters: the portal coupling $\lambda_{\rm hs}$, the quartic $s$ self-coupling $\lambda_{\rm s}$, and the $T=0$ mass of $s$, $m_{\rm s}^2 = \mu_{\rm s}^2 + \lambda_{\rm hs} v^2/2$. 

We assume the $Z_2$ symmetry only for simplicity and, assuming that the $Z_2$ symmetry in the underlying model is broken, neglect all constraints which would be present if $s$ was dark matter \cite{Cline:2013gha}. However, the constraint arising from the Higgs invisible decay must be taken into account. For the range of portal couplings we are considering, $\lambda_{\rm hs}\gsim0.1$, the constraint from the Higgs invisible decay excludes the region $m_{\rm s}<m_{\rm h}/2$.

As in Ref.~\cite{Cline:2012hg} we assume that the necessary $CP$ violation for baryogenesis arises from a dimension-6 operator modifying the top quark mass,
\be
\label{dim6}
y_t \bar Q_L H \left( 1+c\frac{s^2}{\Lambda^2} \right) t_R + {\rm H.c.}
\ee
Here $c$ is a complex number, and $\Lambda$ is a new physics scale. Alternatively, we could consider a dimension-5 operator $\sim s/\Lambda$, but to be consistent with the $Z_2$ symmetric scalar potential, we choose to study the dimension-6 operator. Obviously, our results would not change qualitatively for dimension-5 operator. One should also note that this operator would contribute on the electric dipole moments of the electron and neutron at two loops only if there was mixing between $h$ and $s$ at $T=0$ \cite{Espinosa:2011eu}; thus, the amount of $CP$ violation arising from \eqref{dim6} in the scenario considered here is not constrained by experiments. 

\section{Phase transition pattern} \label{ptpattern}

In the real scalar singlet extension of the Standard Model a first-order electroweak phase transition can be realized at tree-level: First, the $s$ field obtains a nonzero vacuum expectation value. Then, the potential develops a second minimum at $s=0$ which breaks the electroweak symmetry. Finally, this electroweak breaking minimum becomes the global minimum of the potential, and if there is a potential barrier between the electroweak symmetric minimum at $s\neq 0$ and the electroweak breaking minimum at $s=0$, the electroweak phase transition is of first-order. The potential barrier is provided by a sufficiently large portal coupling $\lambda_{\rm hs}$.

To study the phase transition, we include finite temperature corrections to the leading terms in the scalar potential,\footnote{We neglect one-loop corrections beyond the leading $T^2$ terms. Taking into account the full one-loop potential would only slightly change the value of the singlet couplings for which the following results hold.}
\be \label{finiteTpot}
\mu_{\rm s}(T)^2=\mu_{\rm s}^2+c_{\rm s} T^2,\quad \mu_{\rm h}(T)^2=\mu_{\rm h}^2+c_{\rm h} T^2,
\ee
where
\bea
c_{\rm s} &= \frac{1}{12}(2\lambda_{\rm hs}+3\lambda_{\rm s}), \\
c_{\rm h} &= \frac{1}{48}(9 g_L^2+3g_Y^2+12 y_t^2+24\lambda_{\rm h}+2\lambda_{\rm hs}).
\eea
We neglect the contribution $\delta c_{\rm h} = y_t^2 (s/\Lambda)^4/8$ arising from the dimension-6 operator. We will later validate this by checking that $s^2/\Lambda^2$ is small.

Obviously the described phase transition pattern requires that $\mu_{\rm s}^2 < 0$. Moreover, the $s$ direction has to break before the Higgs direction breaks, and the electroweak breaking minimum has to be the global minimum at $T=0$, which require
\be 
\label{cond1}
\frac{\mu_{\rm s}^4}{c_{\rm s}^2} > \frac{\mu_{\rm h}^4}{c_{\rm h}^2},
\ee 
and
\be 
\label{cond2}
\frac{\mu_{\rm s}^4}{\lambda_{\rm s}} < \frac{\mu_{\rm h}^4}{\lambda_{\rm h}},
\ee 
respectively.

\begin{figure}
\begin{center}
\includegraphics[height=.362\textwidth]{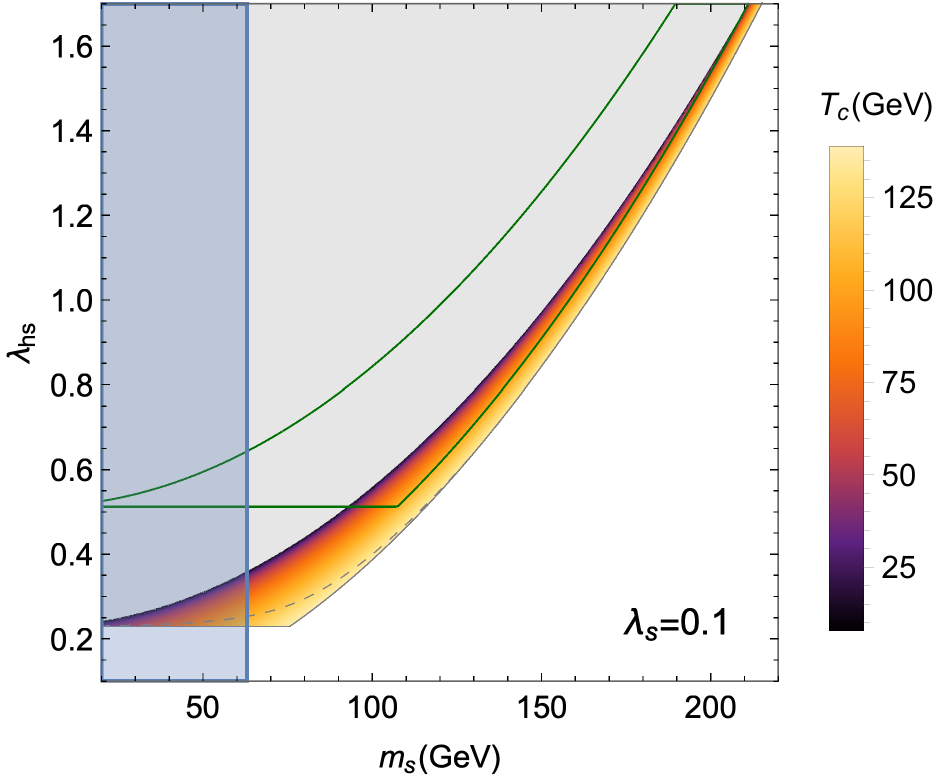}
\caption{Color coding shows the critical temperature in the region where the conditions for the first-order electroweak phase transition are fulfilled for $\lambda_{\rm s} = 0.1$. The dashed line shows the lower limit on $\lambda_{\rm hs}$ requiring that the extremum in the $s$ direction at $T=0.95T_c$ is a minimum. The gray region is excluded for $\lambda_{\rm s} = 0.1$ because there the $T=0$ global minimum of the potential is at $h=0$. In the white region the electroweak phase transition is not of first-order. The green contour marks off the region where the transition is of first-order for $\lambda_{\rm s} = 0.5$. The blue shaded region is excluded by the Higgs invisible decay.}
\label{Tc}
\end{center}
\end{figure}

The critical temperature $T_c$ at which the two minima are equally deep is given by
\be
\label{criticalT}
T_c^2 = \frac{\lambda_{\rm h} c_{\rm s}\mu_{\rm s}^2 - \lambda_{\rm s} c_{\rm h}\mu_{\rm h}^2 - \sqrt{\lambda_{\rm h} \lambda_{\rm s}}|c_{\rm s}\mu_{\rm h}^2 -c_{\rm h}\mu_{\rm s}^2|}{\lambda_{\rm s} c_{\rm h}^2 - \lambda_{\rm h} c_{\rm s}^2}.
\ee
For a first-order electroweak phase transition we must require that the electroweak symmetric extremum is a minimum when the transition occurs. At $T_c$ the condition reads 
\be
\label{cond3}
\lambda_{\rm hs} > 2\sqrt{\lambda_{\rm h}\lambda_{\rm s}},
\ee 
and below $T_c$ the condition becomes more constraining,
\be
\lambda_{\rm hs} (\mu_{\rm s}^2 + c_{\rm s} T^2) < 2 \lambda_{\rm s} (\mu_{\rm h}^2 + c_{\rm h} T^2).
\ee 
The region where the conditions \eqref{cond1}, \eqref{cond2} and \eqref{cond3} are fulfilled is shown in Fig.~\ref{Tc}.

\section{Bubble nucleation and expansion} \label{nucleation}
A first-order phase transition proceeds via nucleation of bubbles of the new phase, which expand and eventually fill the Universe. The bubble nucleation probability per unit time and volume is given by \cite{Linde:1981zj}
\be \label{bubbleprob}
\Gamma \sim T^4\left(\frac{S_3}{2\pi T}\right)^{3/2}\exp\left( -\frac{S_3}{T} \right),
\ee
where
\be \label{fullS3}
S_3 = 4\pi \int r^2 {\rm d}r \left(\frac{1}{2}\left(\frac{{\rm d}h}{{\rm d}r}\right)^2 + \frac{1}{2}\left(\frac{{\rm d}s}{{\rm d}r}\right)^2 + \tilde V \right)
\ee
is the three-dimensional Euclidean action for an O(3)-symmetric bubble corresponding to the path in the field space which minimizes $S_3$. Here $h$ denotes the real part of the neutral component of $H$, and $\tilde V$ is the $Z_2$ symmetric scalar potential \eqref{scalarpotential} including temperature corrections \eqref{finiteTpot} and normalized such that outside the bubble at $r\to\infty$ the potential energy is zero. 

The bubble nucleation temperature $T_{\rm n}$ is defined as the temperature at which the probability of creating at least one bubble per horizon volume is of order one. This condition can be written as \cite{Grojean:2006bp}
\be
\frac{S_3}{T_{\rm n}} \approx -4\log\left(\frac{T_{\rm n}}{M_{\rm Planck}}\right).
\ee

\begin{figure}[b]
\begin{center}
\includegraphics[height=.362\textwidth]{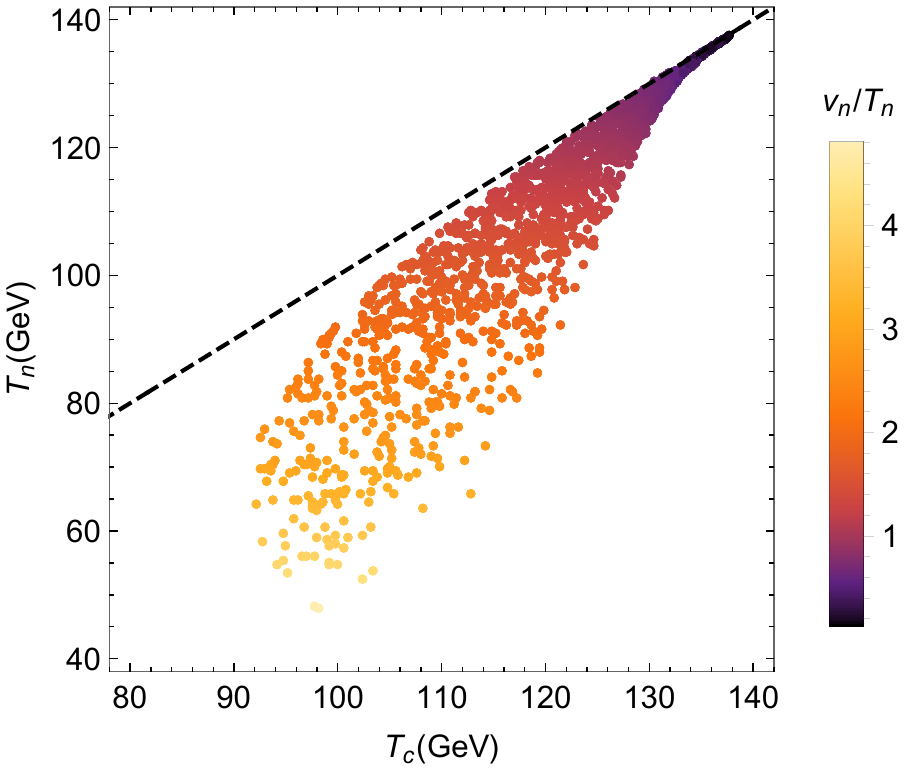}
\caption{The bubble nucleation temperature $T_{\rm n}$ as a function of the critical temperature $T_{\rm c}$ for the points from the scan with $\lambda_{\rm s} = 0.1$. Color coding shows the strength of the transition, $v_{\rm n}/T_{\rm n}$. The dashed line corresponds to $T_{\rm n} = T_{\rm c}$. All points are allowed by the Higgs invisible decay.}
\label{Tn}
\end{center}
\end{figure}

For simplicity, and to speed up numerical calculations, we do not calculate the path which minimizes the full action $S_3$, but we use the path which minimizes the potential energy. It has been checked that typically this approximation works reasonably well \cite{Chung:2011it}. We write the fields as 
\be
h = x \cos\theta \;,\quad s = x \sin\theta \;,
\ee
and for each value of $\theta$, we find the value of $x$ which minimizes the potential energy. Then, knowing the path $x(\theta)$  which minimizes the potential energy, we solve the equation of motion for $\theta$, 
\be
\frac{{\rm d}^2\theta}{{\rm d}r^2} + \frac{2}{r}\frac{{\rm d}\theta}{{\rm d}r} = \frac{1}{x^2} \frac{{\rm d}\tilde V}{{\rm d}\theta} - \frac{1}{x} \frac{{\rm d}x}{{\rm d}\theta} \left(\frac{{\rm d}\theta}{{\rm d}r}\right)^2,
\ee
to find the bubble wall shape as a function of $r$.

\begin{figure*}
\begin{center}
\includegraphics[height=.29\textwidth]{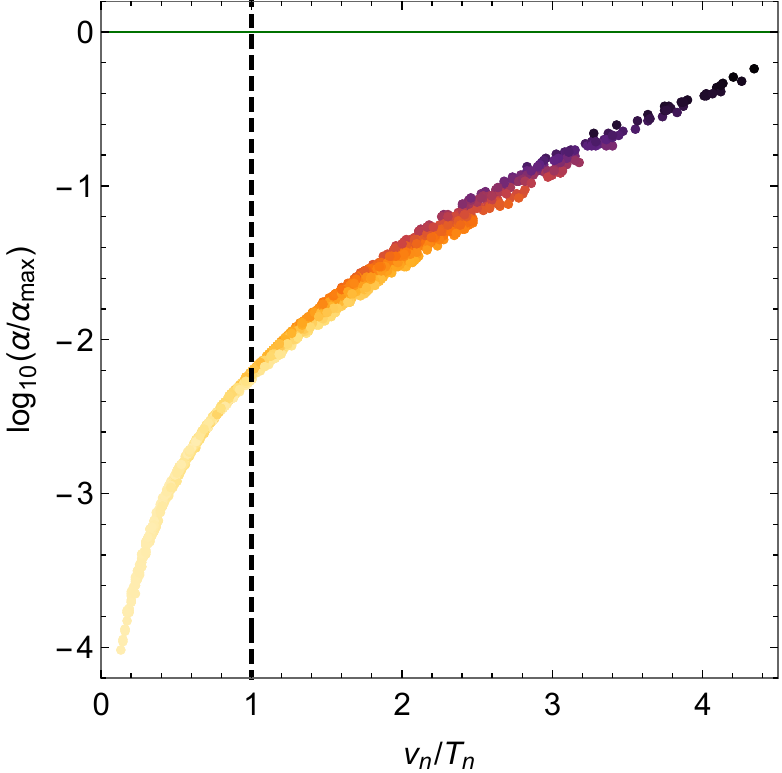} \hspace{.6mm}
\includegraphics[height=.29\textwidth]{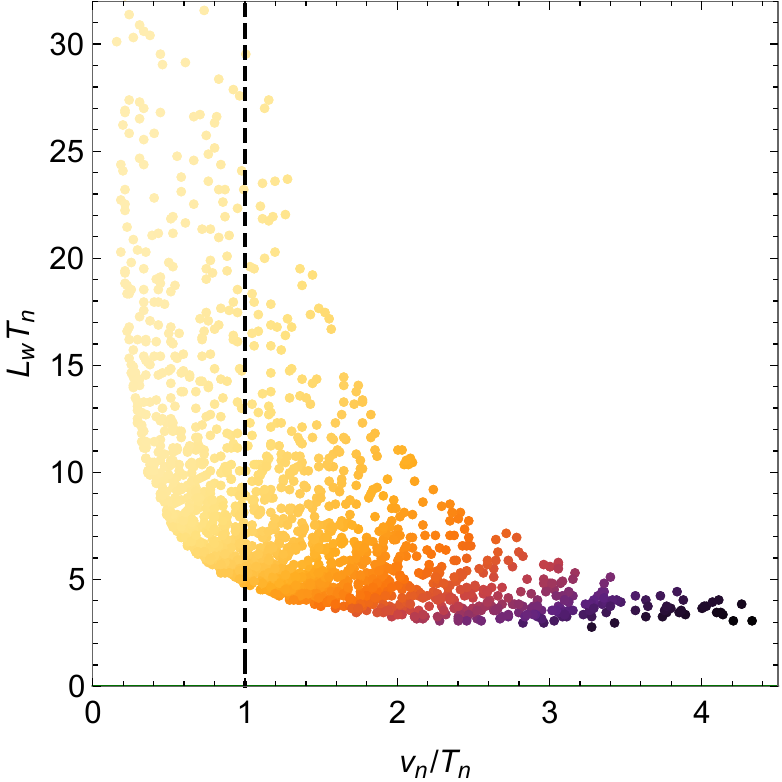} \hspace{.6mm}
\includegraphics[height=.29\textwidth]{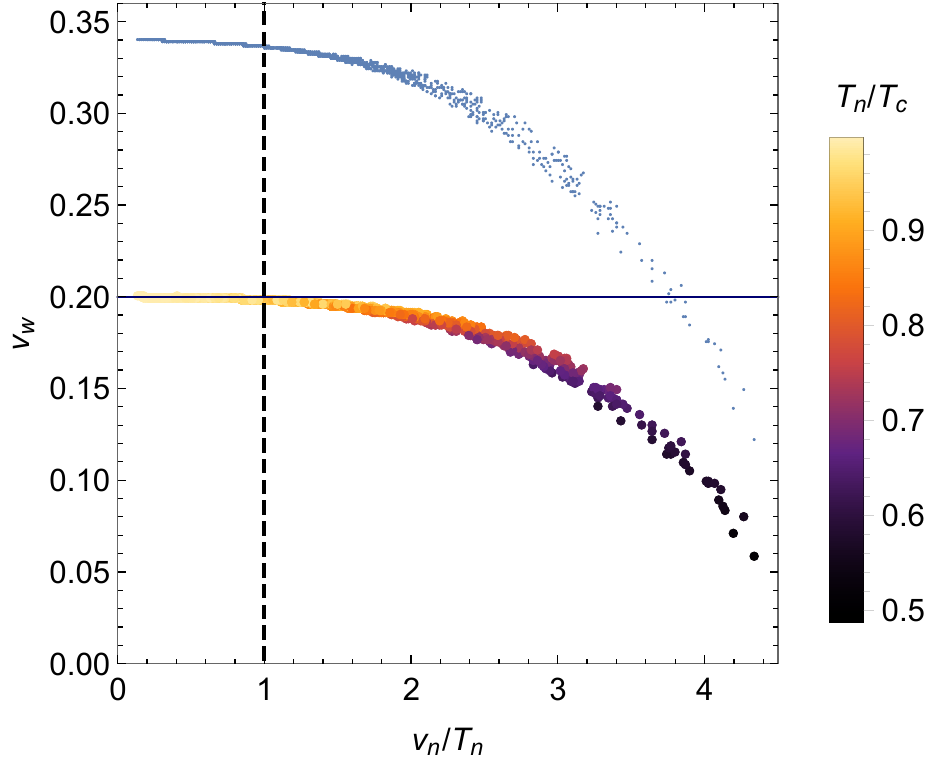}
\caption{Points from the scan with $\lambda_{\rm s}=0.1$. Color coding shows the ratio of the bubble nucleation temperature $T_{\rm n}$ and the critical temperature $T_{\rm c}$. Here $L_{\rm w}$ denotes the bubble wall width, and $v_{\rm w}$ the relative velocity between the bubble wall and the plasma just in front of the wall. To the right of the vertical dashed lines, the transition is sufficiently strong to avoid baryon washout. The blue line in the right panel shows $v_{\rm w} = \xi_{\rm w} = 0.2$, and the blue points show the value of $v_{\rm w}$ for $\xi_{\rm w} = 0.34$.}
\label{hydr}
\end{center}
\end{figure*}

We perform a scan of the parameter space with fixed $\lambda_{\rm s} = 0.1$. We consider only values of $\lambda_{\rm hs}$ and $m_{\rm s}$ which give a first-order electroweak phase transition, e.g. corresponding to the region in the $(m_{\rm s},\lambda_{\rm hs})$ plane shown in Fig.~\ref{Tc}. In Fig.~\ref{Tn} the nucleation temperature is shown for the scanned points. The color coding shows the value of $v_{\rm n}/T_{\rm n}$, where $v_{\rm n}$ denotes the Higgs field expectation value at $T_{\rm n}$ Though all results in this paper are shown only for $\lambda_{\rm s}=0.1$, we have checked that they do not change qualitatively for different values of $\lambda_{\rm s}$.

If the friction force exerted by the plasma on the bubble wall becomes sufficiently large, the bubble wall will quickly reach a constant terminal velocity $\xi_{\rm w} < 1$. Calculating the friction which determines the bubble wall velocity is out of the scope of this work. Instead, we fix $\xi_{\rm w} = 0.2$ which is in agreement with the results from Ref.~\cite{Kozaczuk:2015owa} at $v_{\rm n}/T_{\rm n}\sim 1.1$. For large $v_{\rm n}/T_{\rm n}$ the bubble wall velocity may be significantly larger; thus, in the following sections we will also study how our results would change for different values of $\xi_{\rm w}$.

We accept only the points for which deflagration solutions, necessary for electroweak baryogenesis, exist, e.g. \cite{Espinosa:2010hh}
\be \label{deflcond}
\alpha < \frac{1}{3}(1-\xi_{\rm w})^{-13/10} = \alpha_{\rm max}.
\ee 
Here $\alpha$ is the ratio of released vacuum energy in the transition to that of the radiation bath at $T_{\rm n}$, 
\be
\alpha = \frac{1}{\rho_{\rm \gamma}} \left(\Delta V - \frac{T_{\rm n}}{4} \Delta \frac{{\rm d} V}{{\rm d} T} \right).
\ee
In the left panel of Fig.~\ref{hydr} the ratio $\alpha/\alpha_{\rm max}$ is shown for the scanned points which give $\alpha/\alpha_{\rm max}<1$. We note that even though deflagration solutions exist, it is not guaranteed that they are realized if runaway (or detonation) solutions are also possible. However, we don't find any points that satisfy the criterion $\alpha>\alpha_{\infty}$ \cite{Bodeker:2009qy,Espinosa:2010hh} for the runaway solutions.

\section{Electroweak baryogenesis} \label{ewbg}
The baryogenesis in the model relies on spatially varying complex top quark mass, given by the dimension-6 operator \eqref{dim6}, over the bubble wall. The top quark mass as a function of $z$, which measures the distance from the bubble wall, is given by
\be
m_t(z) = \frac{y_t}{\sqrt{2}} h(z) \left(1+c \frac{s(z)^2}{\Lambda^2}\right).
\ee
We assume that the bubble wall profile is of the form
\bea
&h(z) = \frac{v_{\rm n}}{2} \left(1+\tanh\left(\frac{z}{L_{\rm w}}\right) \right), \\
&s(z) = \frac{w_{\rm n}}{2} \left(1-\tanh\left(\frac{z}{L_{\rm w}}\right) \right),
\eea
where $v_{\rm n}$ and $w_{\rm n}$ are the expectation values of $h$ in broken phase and $s$ in symmetric phase, respectively, at the bubble nucleation temperature $T_{\rm n}$. 

For the bubble wall width we use a very simple estimate \cite{Huber:2013kj}
\be \label{lweq}
L_{\rm w}^2 = \frac{v_{\rm n}^2}{8 V_{\rm b}}, 
\ee
where $V_{\rm b}$ is the height of the potential barrier between the two minima  at $T_{\rm n}$. The bubble wall widths for the scanned points are shown in Fig.~\ref{hydr}. We will later study how the produced baryon-to-entropy ratio changes as a function of $L_{\rm w}$. 

The complex phase of the top quark mass induces a chiral force at the bubble wall region, which causes particles and antiparticles to slow down at different rates. The effect of this force diffuses outside the wall producing a chiral asymmetry in front of the wall. To find the chiral asymmetry which drives the baryon asymmetry production, we solve the chemical potentials $\mu_j(z)$, describing departure from the equilibrium particle densities, for top, antitop and bottom from the transport equations given in Refs.~\cite{Fromme:2006wx,Fromme:2006cm}. From these we construct the left-chiral baryon chemical potential 
\be
\mu_{B_L} = \frac{1}{2}(1+4K_{1,t})\mu_t + \frac{1}{2}(1+4K_{1,b})\mu_b - 2 K_{1,t_c} \mu_{t_c},
\ee
where $K_j$ are thermal averages defined in \cite{Fromme:2006wx}. 

The left-chiral baryon chemical potential enters as a source term to the equation for the baryon number violation rate \cite{Cline:2000nw},
\be 
\label{nBrate}
\dot{n}_B = \frac{3}{2}\Gamma_{\rm sph} \left( 3\mu_{B_L}  T_{\rm n}^2 - \frac{15}{2} n_B \right),
\ee
where the second term in the right-hand side describes baryon number relaxation by the sphaleron processes. Finally the baryon-to-entropy ratio, $\eta_B = n_B/s$, is given by
\be \label{etaBeq}
\eta_B  = \frac{405}{4\pi^2 v_{\rm w} g_{\rm eff} T_{\rm n}} \int_0^\infty {\rm d}z\, \Gamma_{\rm sph} \mu_{B_L} e^{-45 \Gamma_{\rm sph} z/4 v_{\rm w}}.
\ee
For the sphaleron rate we use a formula interpolating between the symmetric and the broken phase~\cite{Arnold:1987zg,Bodeker:1999gx,Cline:2011mm},
\be
\Gamma_{\mathrm{sph}}(z) = \min(10^{-6}T_{\rm n}, 2.4T_{\rm n} e^{-40 h(z)/T_{\rm n}}).
\ee

As emphasized in Ref.~\cite{No:2011fi}, the relevant velocity for baryogenesis is not the bubble wall velocity, but the relative velocity between the bubble wall and the plasma just in front of the wall \cite{Espinosa:2010hh},
\bea \label{vweq}
v_{\rm w} =& \frac{1}{1+\alpha_+}\Bigg( \frac{\xi_{\rm w}}{2} + \frac{1}{6\xi_{\rm w}} \\ &- \sqrt{\left( \frac{\xi_{\rm w}}{2} + \frac{1}{6\xi_{\rm w}} \right)^2 + \alpha_+^2 + \frac{2\alpha_+}{3} - \frac{1}{3}} \Bigg).
\eea
The $\alpha_+$ parameter is given in the Appendix of Ref.~\cite{Espinosa:2010hh}. The velocity $v_{\rm w}$ is much smaller than the bubble wall velocity, especially for very strong transitions. In the right panel of Fig.~\ref{hydr} $v_{\rm w}$ is shown for for the scanned points as a function of $v_{\rm n}/T_{\rm n}$ which characterizes the strength of the transition. For comparison, $v_{\rm w}$ is shown also for $\xi_{\rm w} = 0.34$.

\begin{figure}[b]
\begin{center}
\includegraphics[height=.362\textwidth]{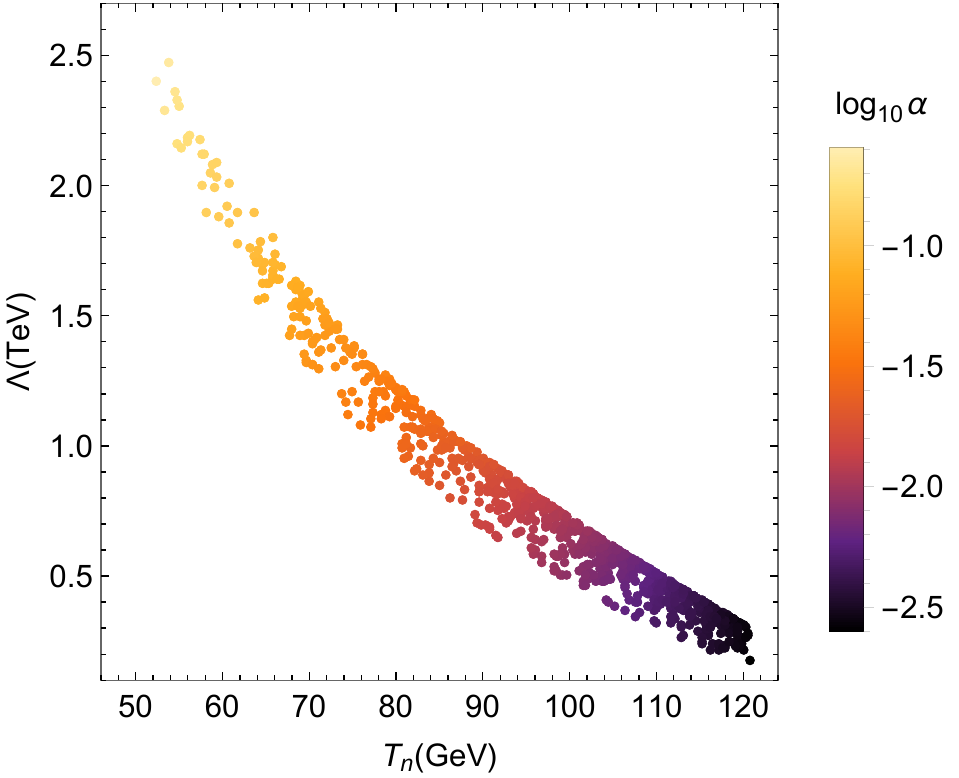}
\caption{The same points as in the right panel of Fig.~\ref{hydr}. The vertical axis shows the new physics scale $\Lambda$ which gives the observed baryon-to-entropy ratio. Color coding shows the ratio of released vacuum energy in the transition to that of the radiation bath at the bubble nucleation temperature.}
\label{lambdaalpha}
\end{center}
\end{figure}

From the scan of the parameter space we take the points for which the electroweak phase transition is sufficiently strong, $v_{\rm n}/T_{\rm n}>1$, to prevent baryon number washout in the electroweak breaking minimum. For these points we perform the baryogenesis calculation. We fix the coupling $c$ to $c = i$. 

First, we notice that the baryon-to-entropy ratio is inversely proportional to the square of the new physics scale $\Lambda$. Hence, we can calculate the baryon-to-entropy ratio for fixed $\Lambda = \Lambda_0$ and then via $\Lambda = \sqrt{\eta_B/\eta_{\rm obs}} \Lambda_0$ we obtain the value of $\Lambda$ which gives the observed baryon-to-entropy ratio $\eta_{\rm obs} = 8.7\times10^{-11}$ \cite{Ade:2015xua}. In Fig.~\ref{lambdaalpha} the values of $\Lambda$ which give the observed baryon-to-entropy ratio are shown for the scanned points. We have also checked that $w_{\rm n}^2/\Lambda^2$ is always small, $w_{\rm n}^2/\Lambda^2\lsim0.1$. Hence, the treatment of the dimension-6 operator is consistent.

From Fig.~\ref{lambdaalpha} we see that $\alpha$ increases as a function of $\Lambda$. The parameter $\alpha$, which measures the vacuum energy released in the transition, increases as a function of $1/T_{\rm n}$. As can be seen from Eq.~\eqref{etaBeq}, the baryon-to-entropy ratio also increases as a function of $1/T_{\rm n}$. Thus, for small $T_{\rm n}$ the new physics scale $\Lambda$ has to be high in order to obtain the observed baryon-to-entropy ratio, because $\eta_B\sim1/\Lambda^2$. This explains the correlation shown in Fig.~\ref{lambdaalpha}: Both $\Lambda$ and $\alpha$ are large for small $T_{\rm n}$. This correlation already points out that the larger $\Lambda$ is, the stronger the gravitational wave signal is, which increases as a function of $\alpha$. We will study in detail the gravitational wave spectrum in the next section.

\begin{figure}
\begin{center}
\includegraphics[height=.318\textwidth]{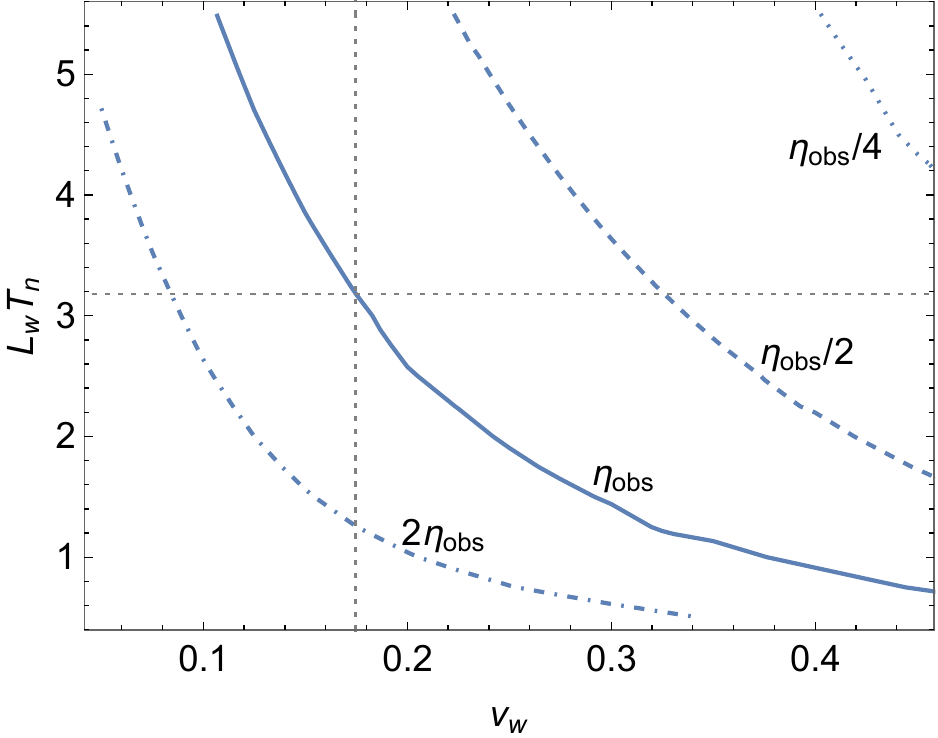}
\caption{Blue lines show the values of the relative velocity between the bubble wall and the plasma just in front of the wall, $v_{\rm w}$, and the bubble wall width $L_{\rm w}$, which give baryon-to-entropy ratios shown in the plot. Here $\lambda_{\rm hs} = 0.554$, $\lambda_{\rm s} = 0.1$, $m_{\rm s} = 114.4$ GeV, and $\Lambda = 1.91$ TeV. Gray dotted lines show the values of $v_{\rm w}$ and $L_{\rm w}$ given by Eqs \eqref{vweq} and \eqref{lweq}.}
\label{vwLw}
\end{center}
\end{figure}

Finally, we show how the baryon-to-entropy ratio depends on $v_{\rm w}$ and $L_{\rm w}$. In Fig.~\ref{vwLw} the baryon-to-entropy ratio is shown in the $(v_{\rm w},L_{\rm w})$ plane for one point from the scan. The baryon-to-entropy ratio decreases as a function of both $v_{\rm w}$ and $L_{\rm w}$. We note that the width of the bubble wall obtained from Eq.~\eqref{lweq} is small for many points, $L_{\rm w} T_{\rm n} \sim 1$. Since the baryogenesis calculation relies on semiclassical analysis which assumes that the bubble wall thickness is much larger than the de Broglie wavelength of particles in the plasma \cite{Cline:2000nw}, the resulting $\eta_B$ for the points with $L_{\rm w} T_{\rm n} \sim 1$ may be inaccurate.

\begin{figure}
\begin{center}
\includegraphics[height=.316\textwidth]{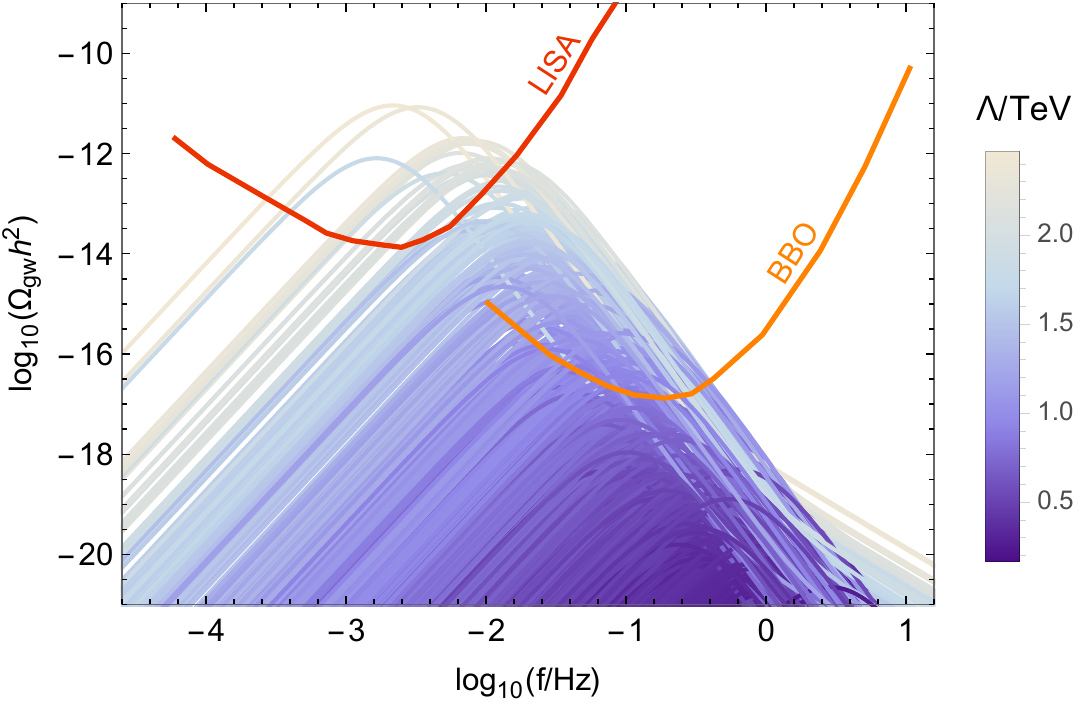}
\caption{Gravitational wave spectra for the same points as in Fig.~\ref{hydr}. Color coding shows the new physics scale which gives the observed baryon-to-entropy ratio. The red and orange curves show the expected sensitivities of LISA and BBO, respectively.}
\label{gwsignal}
\end{center}
\end{figure}

\section{Gravitational wave signal} \label{GW}
The gravitational wave spectrum is determined by the ratio of released vacuum energy in the transition to that of the radiation bath, $\alpha$, the bubble wall velocity $\xi_{\rm w}$, and the inverse duration of the phase transition \cite{Grojean:2006bp},
\be
\beta = H(T_{\rm n}) T_{\rm n} \frac{{\rm d}}{{\rm d}T}\frac{S_3}{T_{\rm n}}.
\ee
For non-runaway bubble walls the gravitational wave signal arises from sound waves and magnetohydrodynamical turbulence in the plasma. We calculate the gravitational wave spectrum, 
\be
\Omega_{\rm gw} h^2(f) = \Omega_{\rm sw}h^2(f) + \Omega_{\rm m}h^2(f),
\ee
following Ref.~\cite{Caprini:2015zlo}. The contributions from sound waves and magnetohydrodynamical turbulence are, respectively, given by
\bea
\Omega_{\rm sw}h^2(f) &= \frac{1.23\times10^{-5}}{g_*^{1/3}} \frac{H}{\beta} \left( \frac{\kappa_{\rm sw} \alpha}{1+\alpha} \right)^2 \xi_{\rm w} S_{\rm sw}(f), \\
\Omega_{\rm m}h^2(f) &= \frac{1.55\times10^{-3}}{g_*^{1/3}} \frac{H}{\beta} \left( \frac{\kappa_{\rm m} \alpha}{1+\alpha} \right)^{\tfrac{3}{2}} \xi_{\rm w} S_{\rm m}(f).
\eea
The functions parametrizing the spectral shape of the gravitational waves read
\bea
S_{\rm sw}(f) &= \left( \frac{f}{f_{\rm sw}} \right)^3 \left( \frac{7}{4+3(f/f_{\rm sw})^2} \right)^{\tfrac{7}{2}} , \\
S_{\rm m}(f) &= \frac{(f/f_{\rm m})^3}{(1+(f/f_{\rm m}))^{\tfrac{11}{3}} (1+8\pi f/h_*)} ,
\eea
with
\be
h_* = 1.65\times10^{-5}\,{\rm Hz}\left( \frac{T_{\rm n}}{100\,{\rm GeV}} \right) \left( \frac{g_*}{100} \right)^{\tfrac{1}{6}}.
\ee
Here $f_{\rm sw}$ and $f_{\rm m}$ are the peak frequencies of each contribution,
\bea
f_{\rm sw} &= \frac{1.9\times10^{-5}{\rm Hz}}{\xi_{\rm w}} \frac{\beta}{H} \left( \frac{T_{\rm n}}{100\,{\rm GeV}} \right) \left( \frac{g_*}{100} \right)^{\tfrac{1}{6}}, \\
f_{\rm m} &= 1.42f_{\rm sw}, \\
\eea
and $\kappa_{\rm sw}$ and $\kappa_{\rm m}$ are the fractions of the released vacuum energy density converted into bulk motion of fluid and magnetohydrodynamical turbulence, respectively. For subsonic bubble walls these read \cite{Espinosa:2010hh}
\bea
\kappa_{\rm sw} &= \frac{c_{\rm s}^{11/5} \kappa_{\rm a} \kappa_{\rm b}}{\left(c_{\rm s}^{11/5}-\xi_{\rm w}^{11/5}\right)\kappa_{\rm b} + \xi_{\rm w}c_{\rm s}^{6/5}\kappa_{\rm a}}, \\
\kappa_{\rm m} &= \epsilon \kappa_{\rm sw},
\eea
where $c_{\rm s} = 1/\sqrt{3}$ is the sound velocity, $\epsilon=0.05$ describes the fraction of bulk motion which is turbulent \cite{Caprini:2015zlo}, and
\bea
\kappa_{\rm a} &= \frac{6.9 \xi_{\rm w}^{6/5}\alpha}{1.36-0.037\sqrt\alpha+\alpha} ,  \\
\kappa_{\rm b} &= \frac{\alpha^{2/5}}{0.017+(0.997+\alpha)^{2/5}} .
\eea

\begin{figure}
\begin{center}
\includegraphics[height=.32\textwidth]{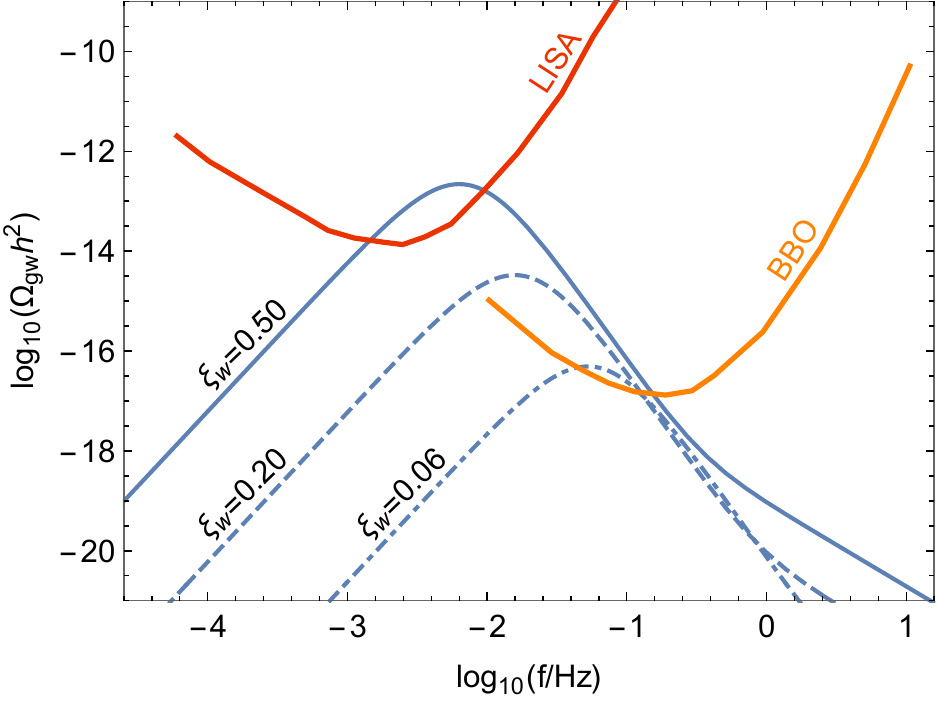}
\caption{Gravitational wave spectrum for $\lambda_{\rm hs} = 0.554$, $\lambda_{\rm s} = 0.1$ and $m_{\rm s} = 114.4$ GeV. The red and orange curves show the expected sensitivities of LISA and BBO, respectively.}
\label{gwxiw}
\end{center}
\end{figure} 

In Fig.~\ref{gwsignal} the gravitational wave spectrum is shown for the points from the scan. Color coding shows the new physics scale $\Lambda$ which gives the observed baryon-to-entropy ratio. Also, the expected sensitivities of LISA and BBO \cite{Thrane:2013oya} are shown. We see that the gravitational wave signal may be well within the reach of the future gravitational wave interferometers and, as expected based on the results presented in the previous section, the strength of the signal increases as a function of $\Lambda$. This is because of the correlation between $\Lambda$ and $\alpha$ shown in Fig.~\ref{lambdaalpha}.

 As mentioned in Sec.~\ref{nucleation} the bubble wall velocity may, in reality for strong transitions, be much larger than the value $\xi_{\rm w} = 0.2$ used for the results shown. Thus, it is interesting to see how the gravitational wave signal depends on the bubble wall velocity. We take the same point as used in Fig.~\ref{vwLw}, and calculate the gravitational wave signal for different bubble wall velocities. In Fig.~\ref{gwxiw} the gravitational wave spectrum is shown for three values of $\xi_{\rm w}$.

\section{Conclusions} \label{conclusions}
We have studied the real scalar singlet extension of the Standard Model where the new scalar field $s$ couples to the Higgs field $h$ via $\lambda_{\rm hs}h^2s^2/4$. For sufficiently large values of the portal coupling $\lambda_{\rm hs}$ the singlet scalar field can induce a strong first-order electroweak phase transition. Also, the $CP$ violation required for baryogenesis is given by the $s$ field via a complex dimension-6 operator, which modifies the top quark mass at $s\neq 0$.

We have shown that if the first-order electroweak phase transition arises from tree-level terms in the potential, the bubble nucleation temperature can be much lower than the critical temperature at which the electroweak symmetric and breaking minima are equally deep. This makes it possible to get a strong gravitational wave signal from the phase transition, since the vacuum energy released in the transition is large. 

We have calculated the baryon-to-entropy ratio by solving the transport equations. Since the baryogenesis does not directly depend on the bubble wall velocity, but the relative velocity between the bubble wall and the plasma just in front of the wall, the observed baryon-to-entropy ratio can be realized at reasonably large values of the new physics scale $\Lambda$. 

Finally, we have calculated the gravitational wave spectrum from the electroweak phase transition. We have compared the gravitational wave signal to the expected sensitivities of LISA and BBO, and shown that these interferometers can test the model. In particular, the parameter space region where the new physics scale $\Lambda$ can be high, is well within the reach of LISA. 

In our analysis we fixed the bubble wall velocity $\xi_{\rm w} = 0.2$. A detailed analysis of the bubble wall dynamics, including a microscopic computation of the friction, is left for future work. The bubble wall velocity is expected to be larger than the value $\xi_{\rm w} = 0.2$ used, especially for large $v_{\rm n}/T_{\rm n}$. Thus, our results give conservative estimates for the gravitational wave signal, as illustrated in Fig.~\ref{gwxiw}. We have also checked, by putting in by hand different increasing behaviors of $\xi_{\rm w}$ as a function of $v_{\rm n}/T_{\rm n}$, that the correlation between $\Lambda$ and the strength of the gravitational wave signal remains.

Also, as indicated by the results of Ref.~\cite{Kozaczuk:2015owa}, the simple estimate used in this work for the bubble wall width may somewhat underestimate the thickness of the wall. Correcting this, and the bubble wall velocity, decreases the baryon-to-entropy ratio. However, on the basis of Fig.~\ref{vwLw} we believe that our results overestimate the new physics scale $\Lambda$ which gives the observed baryon-to-entropy ratio by less than a factor of two. 

\section*{Acknowledgements}
We thank K. Kainulainen and T. Konstandin for discussions. This work was supported by the Estonian Research Council Grant No. IUT23-6 and ERDF Centre of Excellence Project No. TK133.

\bibliography{bubble.bib}

\end{document}